\documentclass[letterpaper,onecolumn]{quantumarticle} 
\pdfoutput=1
\usepackage{amsmath,amsthm,amsfonts,amssymb}
\usepackage{graphicx}

\usepackage{hyperref}
\usepackage{doi}
\usepackage[capitalize,nameinlink]{cleveref}

\usepackage{color}

\newtheorem{lemma}{Lemma}

\usepackage{algorithm}

\DeclareMathAlphabet{\mathpzc}{OT1}{pzc}{m}{it}

\newcommand{\be}{\begin{equation}}
\newcommand{\ee}{\end{equation}}
\newcommand{\cC}{{\cal C}}
\newcommand{\cCT}{{\cal \tilde C}}
\newcommand{\cA}{{\cal A}}
\newcommand{\cQ}{{\cal P}}
\newcommand{\cM}{{\cal M}}
\newcommand{\cB}{{\cal B}}
\newcommand{\CT}{\tilde C}
\newcommand{\kCT}{k_{\tilde C}}
\newcommand{\pCT}{\partial_{\tilde C}}
\newcommand{\cluster}{\kappa}

\DeclareMathOperator{\tc}{TC}

\usepackage[capitalize,nameinlink]{cleveref}

\usepackage{tikz}

\usepackage{url}
\usepackage{algorithm,algorithmic}

\DeclareMathOperator{\im}{Im}

\DeclareMathOperator{\val}{val}

\newcommand{\Z}{\mathbb{Z}}

\newcommand{\HH}{\mathbf H}
\newcommand{\ZZ}{\mathbf Z}
\newcommand{\BB}{\mathbf B}

\begin{document}
\bibliographystyle{unsrturl}

\title{Union-Find Decoders For Homological Product Codes}
\author{Nicolas Delfosse}
\affiliation{Microsoft Quantum and Microsoft Research, Redmond, WA 98052, USA}

\author{Matthew B.~Hastings}

\affiliation{Station Q, Microsoft Research, Santa Barbara, CA 93106-6105, USA}
\affiliation{Microsoft Quantum and Microsoft Research, Redmond, WA 98052, USA}

\begin{abstract}
Homological product codes are a class of codes that can have improved distance while retaining relatively low stabilizer weight.
We show how to build union-find decoders for these product codes, by combining a union-find decoder for one of the codes in the product with a brute force decoder for the other code in the product.  We apply this construction to the specific case of the product of a surface code with a small code such as a $[[4,2,2]]$ code, which we call an {\it augmented surface code}.
The distance of the augmented surface code is the product of the distance of the surface code with that of the small code, and
the union-find decoder, with slight modifications, can decode errors up to half the distance.
We present numerical simulations, showing that while the pseudo-thresholds of these augmented codes are lower than that of the surface code, the low noise performance is improved.
\end{abstract}
\maketitle
The homological product provides a general tool to construct a new CSS quantum code out of two smaller CSS quantum codes.  The resulting codes are called homological product codes\cite{freedman2013quantum,bravyi2014homological}.  One application of this product has been to construct quantum codes with linear distance and rate and with stabilizers whose weight scales only as the square-root of the number of qubits\cite{bravyi2014homological}.  Other applications include weight balancing\cite{hastings2017weight,evra2020decodable} and the construction of some novel code families\cite{audoux2015tensor}.
A special case of the homological product is the hypergraph product\cite{tillich2013quantum}, which has been applied to construct quantum LDPC codes of linear rate and square-root distance.

Any application of a code necessitates a decoder. 
Hypergraph product codes have efficient decoders\cite{leverrier2015quantum,Fawzi_2018}.  
Other decoders for homological product codes considered include an extension of belief propagation\cite{panteleev2019degenerate}, and a decoder for a class of ``three-dimensional codes"\cite{quintavalle2020single}.

In this paper, we give a general construction of an efficient decoder for another class of homological product codes.  The decoder is a generalization of the union-find decoder\cite{delfosse2017almost}.  We assume that one of the codes in the product, which we call the {\it large code} has a union-find decoder, while we assume that the other code, which we call the {\it fixed code}, is of some fixed $O(1)$ size so that we can use a brute force decoder on that code.  From these two ingredients, we construct a decoder on the product code.  (As a technical point, to do this in general, we must assume that the fixed code has no redundancies in its checks, as explained later.)  Importantly, the decoder on the product code is only a constant factor slower than the decoder on the large code and so it runs in close to linear time when the large code is a surface code.

We apply this decoder then to the product of the surface code with various small codes (the product was proposed in \cite{bravyi2014homological} and the anyons of the theory were studied in \cite{vrana2019homological}), and give numerical results.

First let us review the union-find algorithm and also let us say what we mean by a union-find decoder in general: if the large code is not a surface code, what properties should the decoder have in order for it to be called a union-find decoder?
For us, a union find decoder for an LDPC code works as follows: it takes as input some syndrome (and possibly also some list of erased qubits).  It constructs a set of {\it clusters}, where each cluster contains a set of qubits and and a set of checks.
These clusters are constructed using some local operations (i.e., local with respect to the distance defined by the checks of the code) in an initialization phase; for example, in the original union-find decoder with no erasures, then each cluster contains one syndrome error.

Then it runs through an iterative process: for each cluster it checks whether the cluster is {\it valid}.  A cluster is valid if there is some error pattern on the bits in the cluster which produces exactly the observed errors on the checks in that cluster and produces no other errors; let us call that error pattern a {\it valid error pattern}.  The decoder then grows any clusters which are invalid, merging them with other clusters if they intersect, and again checks validity, continuing in this fashion until all clusters are valid.  Then, it constructs some valid error pattern for each cluster (arbitrarily, without regard to minimizing weight), which we call the {\it decoding} of that cluster, and it returns the sum of all those error patterns as the decoding of the input syndrome.

The paper is organized as follows.
In \cref{review}, we review the homological product construction in general.  We also give as an example the special case of the homological product of the surface code with a $[[4,2,2]]$ code.  Interestingly, other authors have considered
{\it concatenating} a surface code with the $[[4,2,2]]$ code\cite{criger2016noise}.  We leave the question of a detailed performance comparison between these two classes of codes (homological product vs. concatenated) for future work, but we note that the homological product gives lower weight stabilizers than concatenation does.

In \cref{sddecode}, we give the decoding algorithm that we use for various products of surfaces codes with small codes.
In particular, we give routines to test validity of clusters, to decode clusters, and to grow clusters, and we present numerical
simulations of this algorithm for a variety of different codes.
The correctness of the test for validity and decoder in this section will follow from the general theory given in
in \cref{general} where we consider more general products, where the large code need not be a surface code and also where the fixed code may have some redundancies in its stabilizers.

Finally, in \cref{distance}, we prove distance properties of the homological product under certain general assumptions on the large code; while in some cases\cite{bravyi2014homological}, the distance of the homological product code can be smaller than the product of the distance of the two codes used as input to the product, if one of the codes is a ``topological code" (in a sense explained later), the distance of the product is equal to the product of the distances.
Further, in \cref{sddecode}, we prove that if the fixed code has distance $2$, then our algorithm decodes up to half the distance of the product, and a modification of our algorithm decodes up to half the distance of the product for arbitrary distance of the fixed code.

\section{Review of Homological Product}
\label{review}
In this section, we review the homological product.  The product we use is a ``multiple sector" product rather than the ``single sector" product used in \cite{bravyi2014homological}.
Throughout this paper, we consider CSS codes over qubits, meaning that all the vector spaces that we define are over $\Z_2$; this means that we may ignore signs throughout.

\subsection{Homology and cohomology}
We consider only $\Z_2$-linear chain complexes and $\Z_2$-homology.
A $D$-dimensional {\em chain complex} is a sequence of $\Z_2$-linear spaces $\cC_i$
$$
\{0\} 
\overset{\partial_{D+1}}{\longrightarrow} 
\cC_D 
\overset{\partial_D}{\longrightarrow} 
\cC_{D-1} 
\cdots
\overset{\partial_2}{\longrightarrow} 
\cC_1
\overset{\partial_1}{\longrightarrow} 
\cC_0
\overset{\partial_0}{\longrightarrow} 
\{0\}
$$
equipped with $\Z_2$-linear maps $\partial_i$ called {\em boundary maps} such that 
$
\partial_{i+1} \circ \partial_i = 0
$
for all $i=0, \dots, D$.
When no confusion is possible, we will omit the subscript $i$.

In the present work, all the spaces $\cC_i$ will be finite dimensional. Vectors of $\cC_i$ are called {\em $i$-chains}. We assume that a basis is fixed for each space $\cC_i$ and we refer to the basis elements as {\em $i$-cells}. 
A $i$-chain can be interpreted equivalently as a binary vector or as a subset of $i$-cells.
As a consequence, we can talk about the boundary of a cell or a set of cells.

Two subsets of the chain space play a central role in homology: 
the {\em cycles} space $\ker \partial_i$ denoted $\ZZ_i$, 
and the space of {\em boundaries} or trivial cycles $\im \partial_{i+1}$ that we denote $\BB_i$.
The quotient $H_i = \ZZ_i / \BB_i$ is the {\em homology} group. In our setting, it is a $\Z_2$-linear space.

We can obtain a new chain complex by replacing each boundary map $\partial_i$ by its transposed map $\partial^i$. Identifying each space $\cC_i$ with its dual, we obtain the complex
$$
\{0\} 
\overset{\partial^{-1}}{\longrightarrow} 
\cC_0
\overset{\partial^0}{\longrightarrow} 
\cC_{1} 
\overset{\partial^1}{\longrightarrow} 
\cdots
\cC_{D-1}
\overset{\partial^{D-1}}{\longrightarrow} 
\cC_D
\overset{\partial^{D}}{\longrightarrow} 
\{0\}
$$
The map $\partial^i$ will be called the {\em coboundary map}
and the space $\ZZ^i = \ker \partial^i$ is the space of {\em cocycles}
and $\BB^i = \im \partial^{i-1}$ is the {\em coboundary} space.
The {\em cohomology group} is defined to the be quotient $H^{i} = \ZZ^i / \BB^i$.

\subsection{General Products}
 The product takes as input two codes.
We describe both codes in terms of chain complexes.

The large code will be defined by some chain complex $\cB$ defined by a sequence of vector spaces $\cB_j$ indexed by some integer $j$ and a boundary map $\partial_B$ from $\cB_{j}$ to $\cB_{j-1}$ such that $\partial_B^2=0$.
For some given $q$, we associate the qubits of the large code with $q$-cells, and we associate $X$- and $Z$-checks of the code with $(q-1)$-cells and $(q+1)$-cells, respectively, with the boundary operator $\partial$ defining the checks of the code in the usual way (original references include \cite{kitaev2003fault,freedman2001projective,bombin2007homological}; see \cite{bravyi2014homological} for a review).
For use later, use $L$ to denote the collection of all cells.
The chain complex defining the large code may be obtained from some cellulation of a manifold, in which case we call it a {\it topological code}, but it need not be.

The fixed code $C$ is defined by some chain complex $\cC$, with three vector spaces $\cC_0,\cC_1,\cC_2$, with boundary operator $\partial_C$ which is a map from $\cC_2$ to $\cC_1$ and from $\cC_1$ to $\cC_0$ such that $\partial_C^2=0$.  Qubits are associated with $1$-cells of $\cC$ and $X$-checks and $Z$-checks are associated with $C_0$ and $C_2$ respectively. Assume that $C$ is an $[[n_C,k_C,d_C]]$ code with $n_X$ distinct $X$ checks and $n_Z$ distinct $Z$ checks.  Assume further that there are no redundancies in the stabilizers of $C$ (i.e., no nontrivial product of stabilizers is the identity), so $n_X+n_Z+k_C=n_C$.  Hence, the zeroth and second homology of $\cC$ vanish.

The homological product code is given by taking the product of the complex defining the large code with the complex defining $C$, and then defining a code from that complex.
The product complex, which we denote $\cA$, has spaces $\cA_0,\cA_1,\ldots$, with
$$\cA_i=\oplus_{j=0}^i (\cB_j \otimes \cC_{i-j}),$$
and the product complex has boundary operator
$$\partial=\partial_B \otimes I + I \otimes \partial_C.$$
Note that if the code is over qudits rather than qubits, there is a slightly more complicated sign structure for the boundary operator.

In the homological product code, qubits are identified with basis vectors of the space $\cA_{q+1}$, i.e.,
$$(\cB_{q-1}\otimes \cC_2) \oplus (\cB_q \otimes \cC_1) \oplus (\cB_{q+1} \otimes \cC_0).$$
Thus, for every $(q+1)$-cell of $L$, there are $n_X$ qubits, for every $q$-cell of $L$ there are $n_C$ qubits, and for every $(q-1)$-cell of $L$ there are $n_Z$ qubits.

The product code is an $[[n,k,d]]$ code where,
by the Kunneth formula, if the large code is an $[[n_B,k_B,d_B]]$ code, then given our assumption on the zeroth and second homology of $\cC$, $$k=k_B k_C.$$

Logical $Z$ operators of the product are given by  $(q+1)$-th homology classes (a particular choice of logical operator corresponds to a representative) and logical $X$ operators are given by $(q+1)$-th cohomology classes.

For a surface code on a square lattice, the number of $2$-cells is roughly $n_B/2$, depending on boundary conditions, and similarly for the number of $0$-cells.
So, in this case $$n\approx \frac{n_B (n_X+n_Z)}{2}+n_B n_C.$$

$X$-stabilizers of the product code are associated with basis vectors of $\cA_q=$
$$(\cB_{q-2} \otimes \cC_2) \oplus (\cB_{q-1}\otimes \cC_1) \oplus (\cB_q \otimes \cC_0),$$
so that for every $(q-2)$-cell there are $n_Z$ different stabilizers,
for every $(q-1)$-cell there are $k_C$ different stabilizers and for every $q$-cell that are $n_X$ different stabilizers.
Note that in many cases (for example, the surface code with $q=1$ or some LDPC codes), the first term
$(\cB_{q-2} \otimes \cC_2)$ is absent.

The ``cells" of the product complex will be basis vectors of spaces $\cA_q$, and a cell $e$ of the product complex will be {\it associated} with some cell $f$ of $L$ if the basis vector corresponding to $e$ is some product of the basis vector corresponding to $f$ with some basis vector of a cell in the fixed code.
Given any vector $v$ in this space $\cA_q$, and any $(q-1)$-cell $e\in L$, we define the {\it vector of coefficients} of $v$ on $e$ on that cell in the obvious way: it is given by the $n_C$ distinct coefficients of $v$ for cells of the product complex which are associated to $e$.

\subsection{Simple Example}
\begin{figure}
\centering
(a) 
\includegraphics[scale=.45]{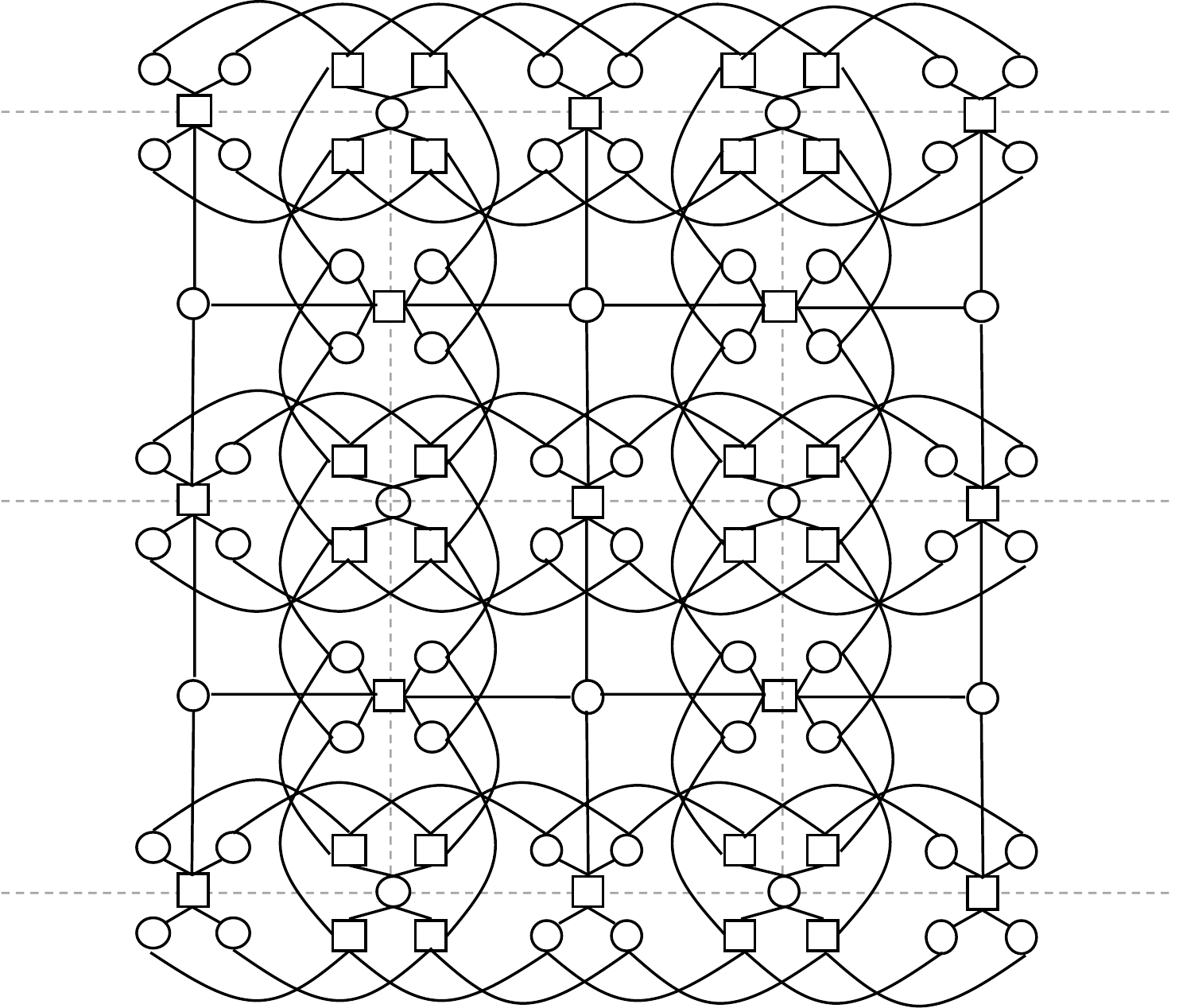}

\bigskip
(b) 
\includegraphics[scale=.45]{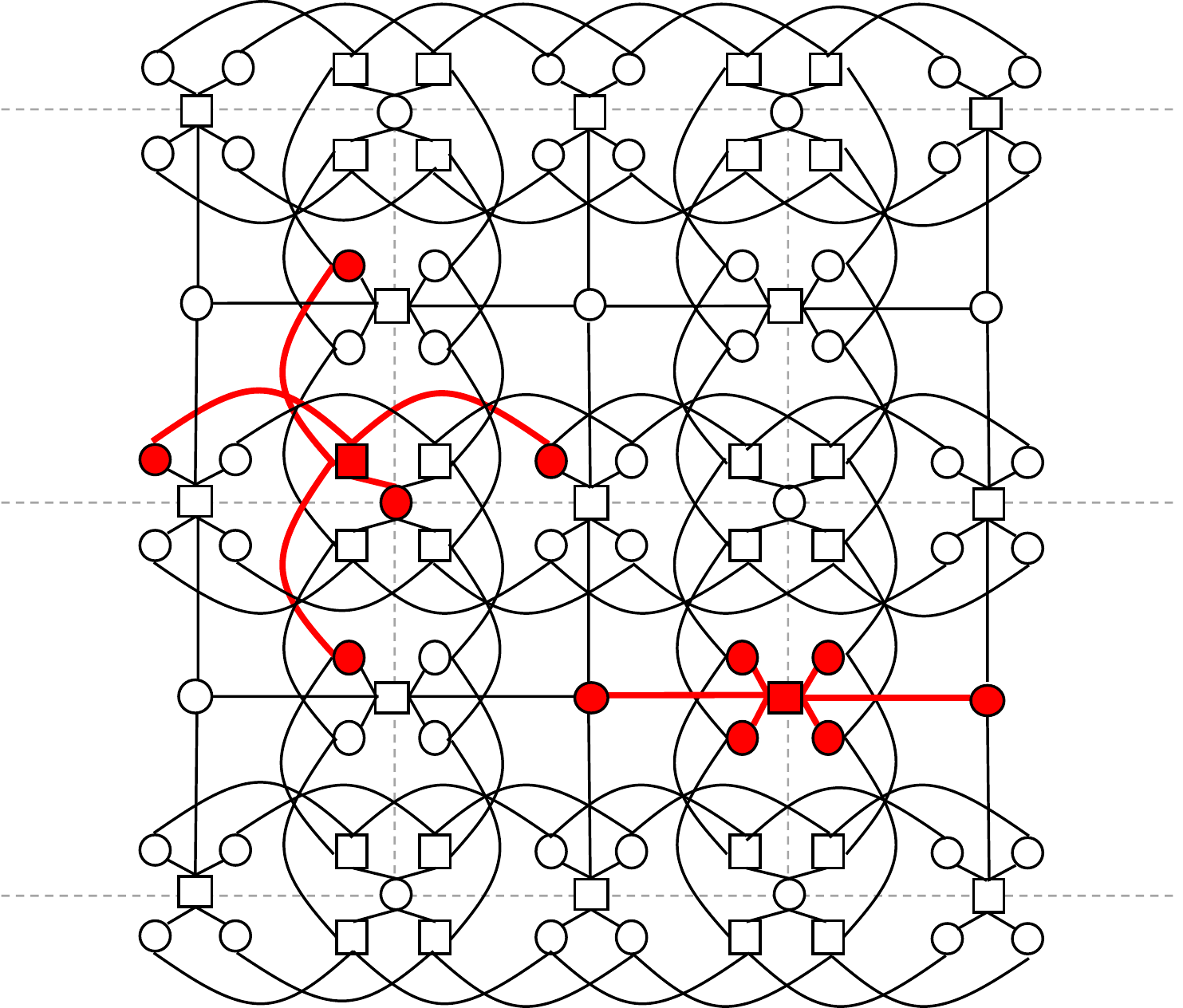}
(c) 
\includegraphics[scale=.45]{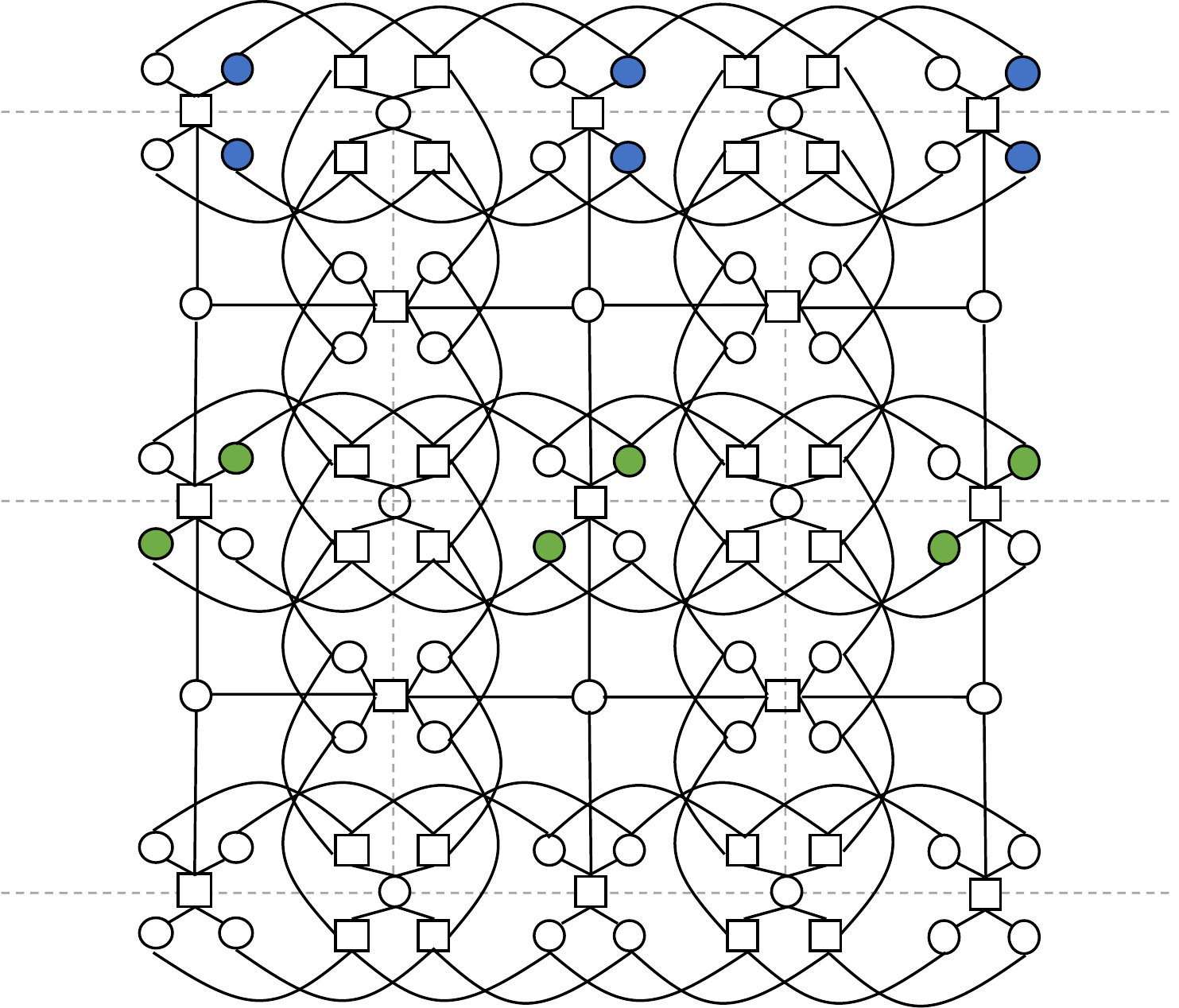}
\caption{
(a) Homological product of a distance-three surface code with a $[[4, 2, 2]]$ code. Dashed lines represent the original lattice of the surface code. 
Qubit are represented by circles and square represent the $X$-checks. A check acts on its neighboring qubits. One can obtain a similar representation of $Z$-checks by considering the dual lattice.
(b) Two checks and their support. All the checks are local and have weight at most 6. 
(c) Two distinct $Z$ logical operators for the two logical qubits of the product code.
}
\label{fig:surface_code_x_422}
\end{figure}

As an example of this formalism, we give perhaps the simplest augmented surface code, the homological product of a surface code on a square lattice with a $[[4,2,2]]$ code with checks $XXXX$ and $ZZZZ$. Figure~\ref{fig:surface_code_x_422} shows the layout of the qubits are checks of the augmented surface code.

This code has $4$ qubits per edge, one qubit per vertex, and one qubit per plaquette.  It has one $X$ stabilizer on each edge $e$ given by
$$(\prod_{a=1}^4 X_e^a) (\prod_{p, e\in \partial p} X_p),$$
where $X_e^a$ for $a=1,\ldots,4$ are Pauli operators on the four qubits on edge $e$, the product is over plaquettes $p$ such that $e$ is in the boundary of $p$, and $X_p$ is a Pauli operator on the qubit on plaquette $p$.
This stabilizer is weight $6$.
It has four $X$ stabilizers on each vertex $v$ given by (for $a=1,\ldots,4$)
$$(\prod_{e, v\in \partial e} X_e^a) X_v,$$
where $X_v$ is a Pauli operator on the qubit on vertex $v$.
This stabilizer is weight $5$.
There is also one $Z$ stabilizer on each edge $e$ given by
$$(\prod_{a=1}^4 Z_e^a) (\prod_{v\in \partial e} Z_v),$$
and four $Z$ stabilizers on each plaquette $p$ given by
$$(\prod_{e\in \partial p} Z_e^a) Z_p.$$

\begin{figure}
\centering
\includegraphics[scale=.5]{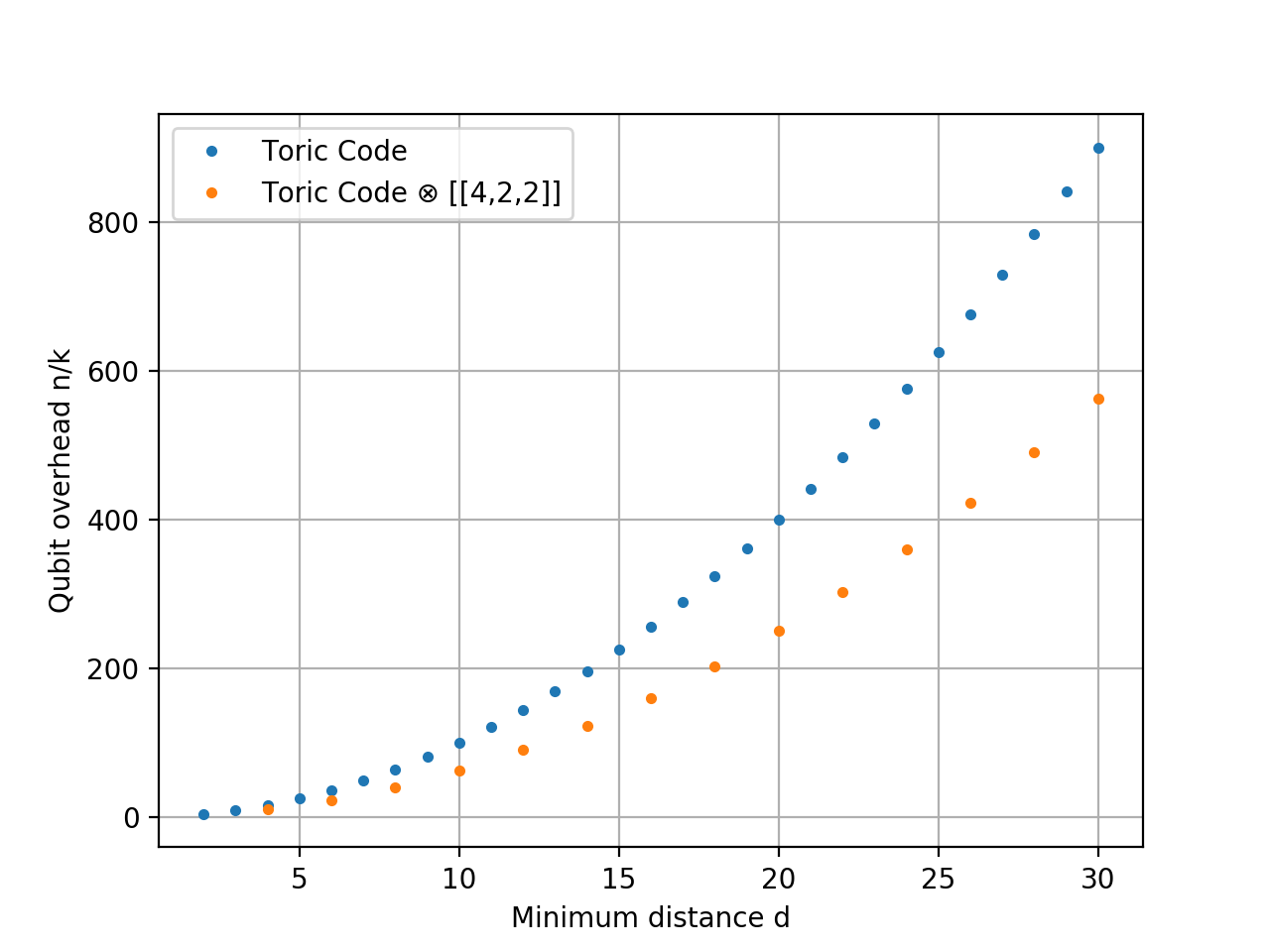}
\caption{
Qubit overhead $n/k$ as a function of the minimum distance for the toric code
and augmented toric code. 
The toric code reaches $d=30$ with 900 physical qubits
per logical qubit while the augmented toric code requires only 
562.5 physical qubits per logical qubit saving almost $40\%$ 
of the qubits.
}
\label{fig:distance_vs_overhead}
\end{figure}

For any cellulation, the code has twice as many logical qubits as the surface code does.
Based a $m \times m$ square lattice of a torus one can define a 
$[[2m^2, 2, m]]$ toric code. Its product with $[[4, 2, 2]]$ code provides a
$[[10m^2, 4, 2m]]$ which uses more physical qubits but encodes 
twice has many logical qubits with a doubled minimum distance (Section~\ref{distance}).
Overall, this strategy leads to a significant reduction of the qubit overhead required to
achieve a given minimum distance as shown in Figure~\ref{fig:distance_vs_overhead}.

\section{Decoding Augmented Surface Codes}
\label{sddecode}
We now give our decoding algorithm for the case where the large code is a surface code and the fixed code has no redundancies in its checks.
Since we consider CSS codes, we consider
just the correction of $Z$-errors using $X$-type measurements.
By duality, $X$-type errors can be corrected with the same procedure.
So, for us, the error patterns will be on $2$-cells of the product complex and 
the syndrome of an error will be the set of 1-cells on its boundary.

Before giving a detailed, technical description of decoding, let us review the union-find decoder for the surface code at a high level.  The reader should see \cite{delfosse2017almost} for more details.  The algorithm works by growing {\it clusters} of cells.  Initially, there is exactly one cluster for each error, with the cluster containing just the single cell on which that error occurs.  As the algorithm runs, clusters grows by adding $1$-cells, and when clusters join together, they merge into a single, larger cluster.  A cluster is {\it valid} if it contains an even number of errors, which means that the error pattern on that cluster can be produced just by errors on the edges in that cluster.  Clusters are not valid if they contain an odd number of errors, in which case to produce the observed error pattern, there must be at least one error on an edge leaving the cluster.  This motivates the growth rules: only clusters which are not valid grow while clusters which are valid do not grow (though they can merge with a growing cluster which is not valid).  Eventually, once growth stops and all clusters are valid, the decoder then applies a correction supported just on the clusters.  The name ``union-find" refers to a data structure used to keep track of the growing clusters.  The decoder here is a generalization of the union-find decoder, using in particular a more general way to determine validity of clusters.

\subsection{Error and syndrome}

To define the surface code, consider a finite cellulation $(V, E, F)$ of a closed surface 
with vertex set $V$, edge set $E$ and face set $F$.
The chain complex $\cB$ of the surface code is the chain complex of the cellulation
defined over the spaces $\cB_2 = \Z_2^F$,
$\cB_1 = \Z_2^E$ and $\cB_0 = \Z_2^V$.

The fixed code $C$ is a $[[n_C, k_C]]$ CSS code defined by a pair of binary 
matrices $\HH_X, \HH_Z$ with $n_C$ columns satisfying $\HH_Z^T \cdot \HH_X = 0$. 
The $n_X$ rows of $\HH_X$ correspond to the $X$ type stabilizer generators 
and $n_Z$ rows of $\HH_Z$ define the $Z$ type stabilizer generators of $C$.
Equivalently, the code $C$ can be described by the chain complex
$\cC_2 = \Z_2^{n_Z} \rightarrow \cC_1 = \Z_2^{n_C} \rightarrow \cC_0 = \Z_2^{n_X}$.
The matrices $\HH_X$ and $\HH_Z^T$ are the matrices of the
boundary maps $\cC_1 \rightarrow \cC_0$ and $\cC_2 \rightarrow \cC_1$
respectively. 

Qubits of the augmented surface code are placed on the $2$-cells of the homological product 
$\cA = \cB \otimes \cC$.
A $Z$ error, {\em i.e.} a 2-chain 
$x \in \cA_2 = (\cB_0 \otimes \cC_2) \oplus (\cB_1 \otimes \cC_1) \oplus (\cB_2 \otimes \cC_0)$
can be uniquely written in the standard form
\begin{equation} \label{eqn:error}
x = \sum_{v \in V} v \otimes x(v) +  \sum_{e \in E} e \otimes x(e) +  \sum_{f \in F} f \otimes x(f)
\end{equation}
where $x(v) \in \cC_2, x(e) \in \cC_1$ and $x(f) \in \cC_0$.
The 2-chain associated with trivial vectors $x(v), x(e)$ and $x(f)$ for all the cells of 
the surface is the trivial 2-chain.

The syndrome $s(x)$ of an error $x$ as in \eqref{eqn:error} is obtained by applying the
boundary map of the product complex:
\begin{equation} \label{eqn:syndrome_eval}
s(x) = \sum_{v \in V} v \otimes \HH_Z^T x(v)^T 
+  \sum_{e \in E} e \otimes \HH_X^T x(e)^T + \sum_{e \in E} \partial(e) \otimes x(e) 
+  \sum_{f \in F} \partial(f) \otimes x(f)
\end{equation}
One can write this syndrome using the standard form 
\begin{equation} \label{eqn:syndrome}
s = \sum_{v \in V} v \otimes s(v) +  \sum_{e \in E} e \otimes s(e)
\end{equation}
with $s(v) \in \cC_1$ and $s(e) \in \cC_0$. The first and third terms of \eqref{eqn:syndrome_eval}
contribute to the vectors $s(v)$ and the second and fourth terms contribute to
$s(e)$.

\subsection{Generalization of the Union-Find decoder}

In order to design a decoder for augmented surface codes, we 
use a decoder $D_X: \cC_0 \rightarrow \cC_1$
and a decoder $D_Z^T: \cC_1 \rightarrow \cC_2$
for the linear codes with parity check matrices $\HH_X$ and 
$\HH_Z^T$ respectively. For a small CSS code, these two decoders
can be given as a lookup table. We can assume that they provide
a minimum weight correction.

Pick a family $x_1, \dots, x_{k_C} \in \cC_1$ representing ${k_C}$ independent $X$ logical operators of the fixed CSS code $C$ and let 
$z_1, \dots, z_{k_C} \in \cC_1$ be independent $Z$ logical operators of $C$
such that $(x_i | z_j) = \delta_{i, j} \pmod 2$.

The union-find decoder will grow connected clusters in the surface until
all the clusters can be erased and corrected independently given the syndrome
$s$. 
Such a cluster is said to be a {\em valid cluster}.
To determine if a cluster $\cluster$ is valid, we compute the validity vector 
\begin{align} \label{eqn:validity}
\val(\cluster) = \sum_{v \in \cluster} ( (s(v) | x_1), \dots, (s(v) | x_{k_C}) ) \in \Z_2^{k_C}
\end{align}
A cluster is valid iff its validity vector is trivial. 
In Section~\ref{valid}, we will prove that a valid cluster contains a valid error pattern.

\begin{algorithm}
\caption{Union-Find decoder for augmented surface codes}
\label{algo:uf_decoder}

\begin{algorithmic}[1]
\REQUIRE The syndrome $s(x)$ as in \eqref{eqn:syndrome} of an error $x \in A_2$.\\
\ENSURE An estimation $\tilde x \in A_2$ of $x$.

\STATE Set $\tilde x = 0$ and ${\cal E} = \{ e \in E \ | \ s(e) \neq 0 \}$.
\STATE {\bf Edge cancellation:}
\STATE Run over all edges $e = \{ u, v \} \in E$ and do:
\STATE \hspace{1cm} Compute $\tilde x(e) = D_X(s(e))$.
\STATE \hspace{1cm} Add $e \otimes \tilde x(e)$ to $\tilde x$.
\STATE \hspace{1cm} Add $u \otimes s(e)$ and $v \otimes s(e)$ to $s$.
\STATE {\bf Union-Find growth:} 
\STATE Initialize clusters with a single vertex $\cluster_{v} = \{v\}$.
\STATE Merge clusters connected by an edge of ${\cal E}$.
\STATE While there exists at least one invalid cluster do:
\STATE \hspace{1cm} Select an invalid cluster $\cluster$ with minimum boundary.
\STATE \hspace{1cm} Grow $\cluster$ by one half-edge.
\STATE \hspace{1cm} Update the validity vector of $\cluster$.
\STATE Set ${\cal E}'$ to be the set of all edges fully covered by the grown clusters.
\STATE {\bf Peeling:} 
\STATE Construct a spanning forest ${\cal F}$ of the subgraph ${\cal E}'$ of the surface.
\STATE While ${\cal F} \neq \emptyset$ do:
\STATE \hspace{1cm} Select an edge $e = \{ u, v \}$ of ${\cal F}$ such that $u$ is a leaf.
\STATE \hspace{1cm} Add $e \otimes s(u)$ to $\tilde x$.
\STATE \hspace{1cm} Add $u \otimes s(u)$ and $v \otimes s(u)$ to $s$.
\STATE \hspace{1cm} For $i=1, \dots, k_C$ do:
\STATE \hspace{2cm} If $(s(u)|x_i) = 1$ do:
\STATE \hspace{3cm} Add $e \otimes z_i$ to $\tilde x$.
\STATE \hspace{3cm} Add $u \otimes z_i$ and $v \otimes z_i$ to $s$.
\STATE \hspace{1cm} Remove $e$ from ${\cal F}$
\STATE {\bf Residual Node Correction:} 
\STATE For all $v \in V$ do:
\STATE \hspace{1cm} Compute $\tilde x(v) = D_Z^T(s(v))$.
\STATE \hspace{1cm} Add $v \otimes \tilde x(v)$ to $\tilde x$.
\STATE Return $\tilde x$.
\end{algorithmic}
\end{algorithm}

Algorithm~\ref{algo:uf_decoder} works in four steps. 
First, we eliminate the components of the syndrome on edges.
A syndrome component $e \otimes s(e) \in B_1 \otimes C_0$ is killed 
(moved to the endpoints of $e$) by adding an $Z$ error on 
the 2-cell $e \otimes D_X(s(e)) \in B_1 \otimes C_1$.
We will treat this edge set ${\cal E} \subset E$ as an erasure 
in the surface.

The second step is a growth step similar to the union-find 
decoder. On can erase and decode a cluster iff it is valid.
We apply the same growth procedure as in the union-find decoder
with erasure ${\cal E}$ and with the validity function defined above
based on the syndrome values $s(v)$ on nodes.

Once valid clusters are grown, they are erased and the peeling 
decoder \cite{delfosse2020peeling} is used to identify a correction that explains the 
syndrome inside each cluster. 
This peeling step applies a
correction on the edges of a spanning tree of each cluster.
Over an edge $e = \{u, v\}$, two types of correction are applied.
(i) A correction $e \otimes s(u)$ moves the local syndrome $s(u)$ 
from $u$ to $v$.
(ii) A correction of the form $e \otimes z_i$ to used to cancel the 
validity vector in node $u$.
As a result, after peeling, each node $v$ of the cluster, except the root $v_0$,
supports a trivial syndrome $v \otimes s(v)$ with $s(v) = 0$.
Moreover, the validity vector of each node of the cluster is trivial
because the cluster is valid.

Finally, the residual syndrome $v_0 \otimes s(v_0)$ on each cluster root 
$v_0$ is eliminated using the decoder $D_Z^T$ by applying a correction
$v_0 \otimes D_Z^T(s(v_0))$. This local decoder can be applied
because $s(v_0)$ is a $Z$ stabilizer of $C$. 
Indeed, we will show that $s(v_0)$ is orthogonal with $X$ stabilizers
and $X$ logical operators of $C$.
The root $v_0$ satisfies $\val(v_0) = 0$, which means that $s(v_0)$ 
is orthogonal with all the $X$ logical operators of $C$.
Moreover, since the cluster is valid before peeling, the restriction of the syndrome
to the cluster, is 1-boundary in $\cA$ and this property is preserved after peeling. 
As a consequence the $v_0 \otimes s(v_0)$ is a 1-boundary and it satisfies 
$\partial( v_0 \otimes s(v_0) ) = 0$ (because $\partial \circ \partial = 0$).
This implies $\HH_X s(v_0)^T = 0$, showing that $s(v_0)$ is orthogonal with
all $X$ stabilizers.

One can slightly improve the performance of Algorithm~\ref{algo:uf_decoder} by modifying the growth procedure. This variant of the decoder is described in Appendix~\ref{app:UF_modified}.

\subsection{Numerical results}

In this section, we report the result of numerical simulation of augmented toric codes. 
We consider the product of a rotated toric code with different CSS codes. 
We only simulate the correction of $Z$-error since the codes considered are
self-orthogonal.
To sample from $Z$-errors, we start from the trivial 2-chain and 
we flip each bit of each vector $x(v), x(e)$ and $x(f)$ independently 
with probability $p$.  For simplicity, we do not consider the case of erasure errors.
We assume perfect syndrome extraction circuit. Our numerical simulations are based on the variant of Algorithm~\ref{algo:uf_decoder} proposed in Appendix~\ref{app:UF_modified}.

In Figure~\ref{fig:distance_vs_overhead} we observed that augmented topological codes can achieve the same distance as traditional topological codes with a reduced qubit overhead.
To illustrate further the advantage of augmented codes, we compare numerically the performance of toric codes and their product with a $[[4, 2, 2]]$ code. In both cases, the correction is performed using a union-find decoder.
Denote by $\tc(m)$ the toric code defined on the $m \times m$ lattice. 

\begin{figure}
\centering
\includegraphics[scale=.45]{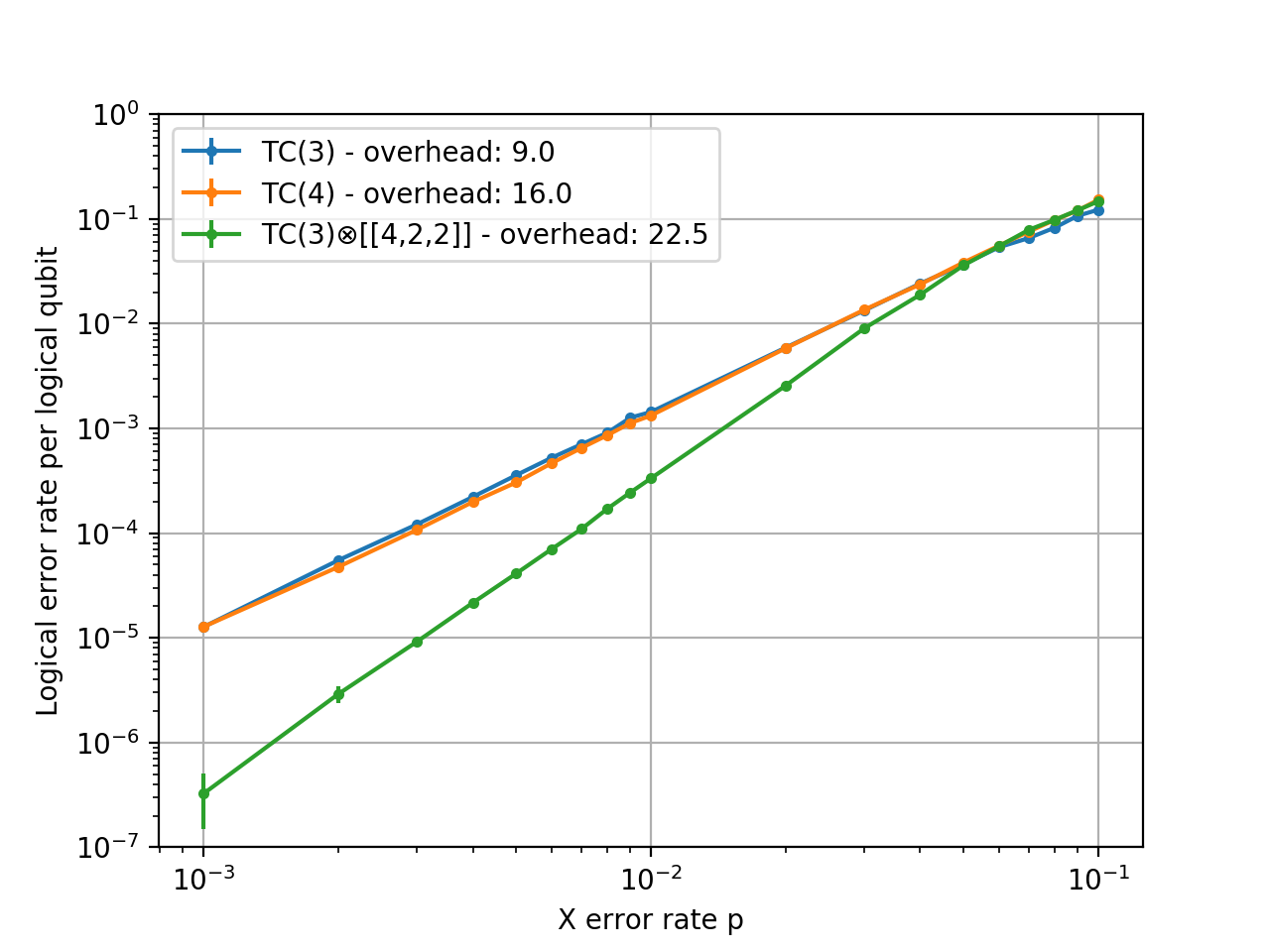}
\includegraphics[scale=.45]{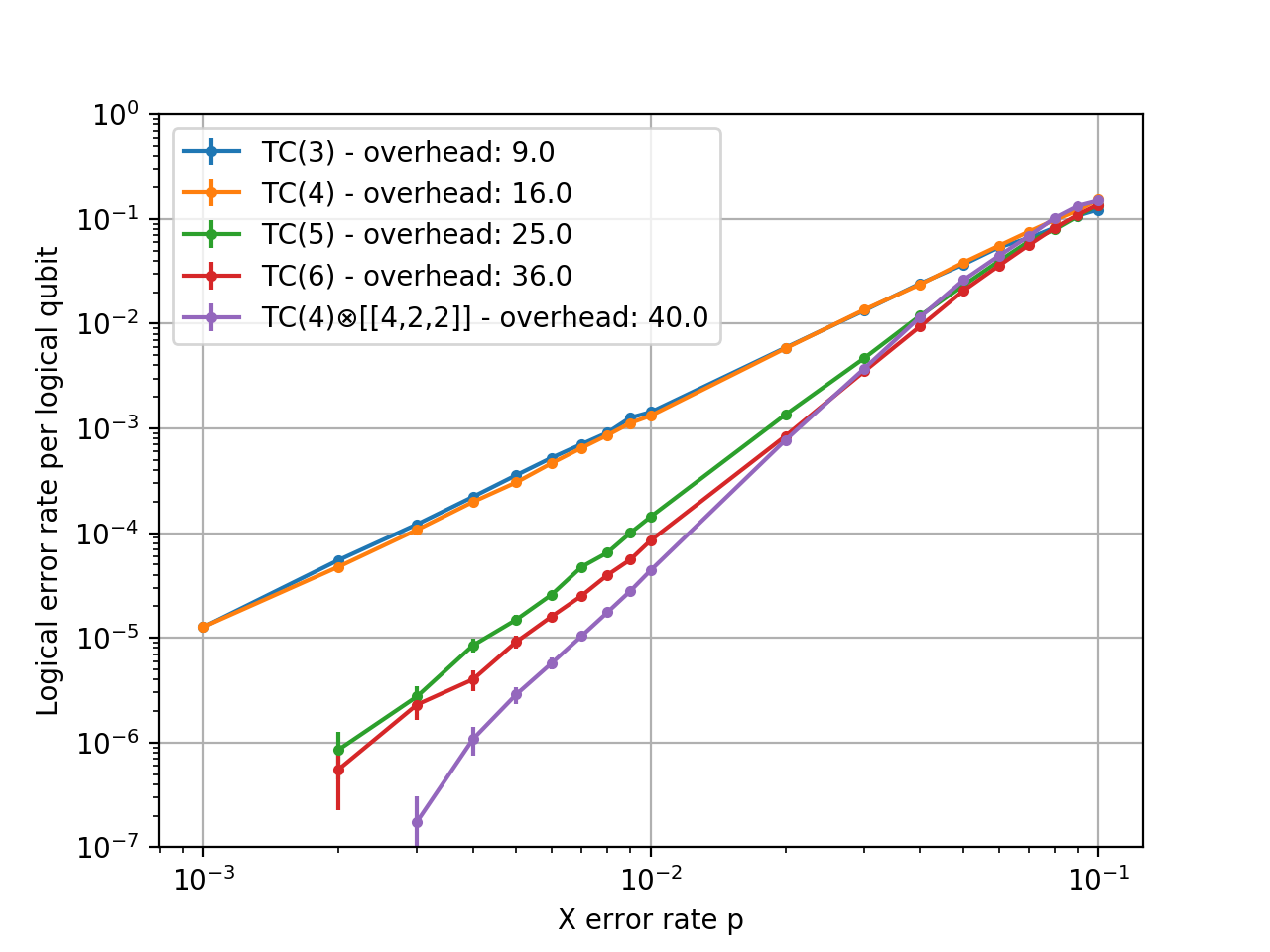}
\includegraphics[scale=.45]{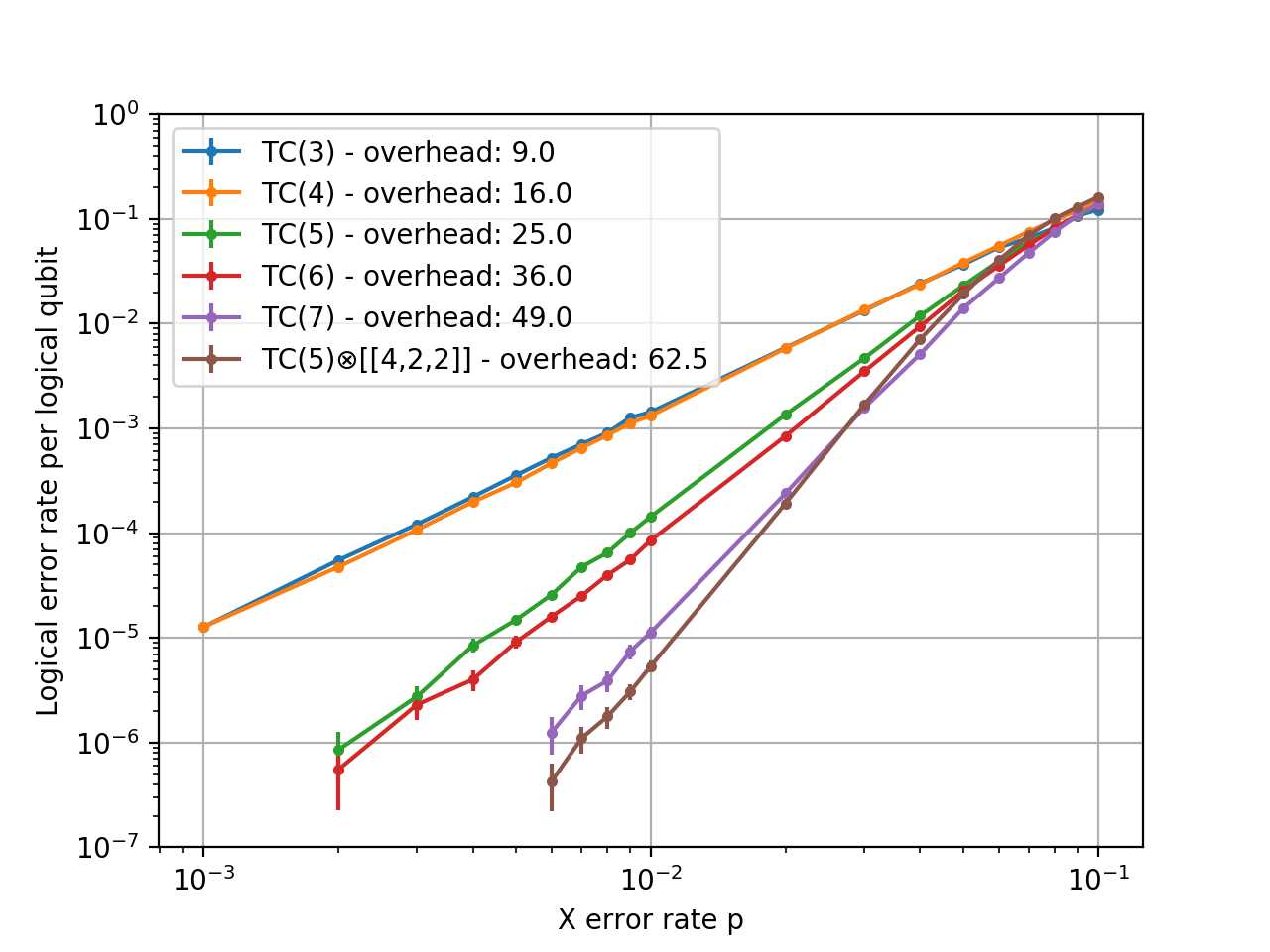}
\includegraphics[scale=.45]{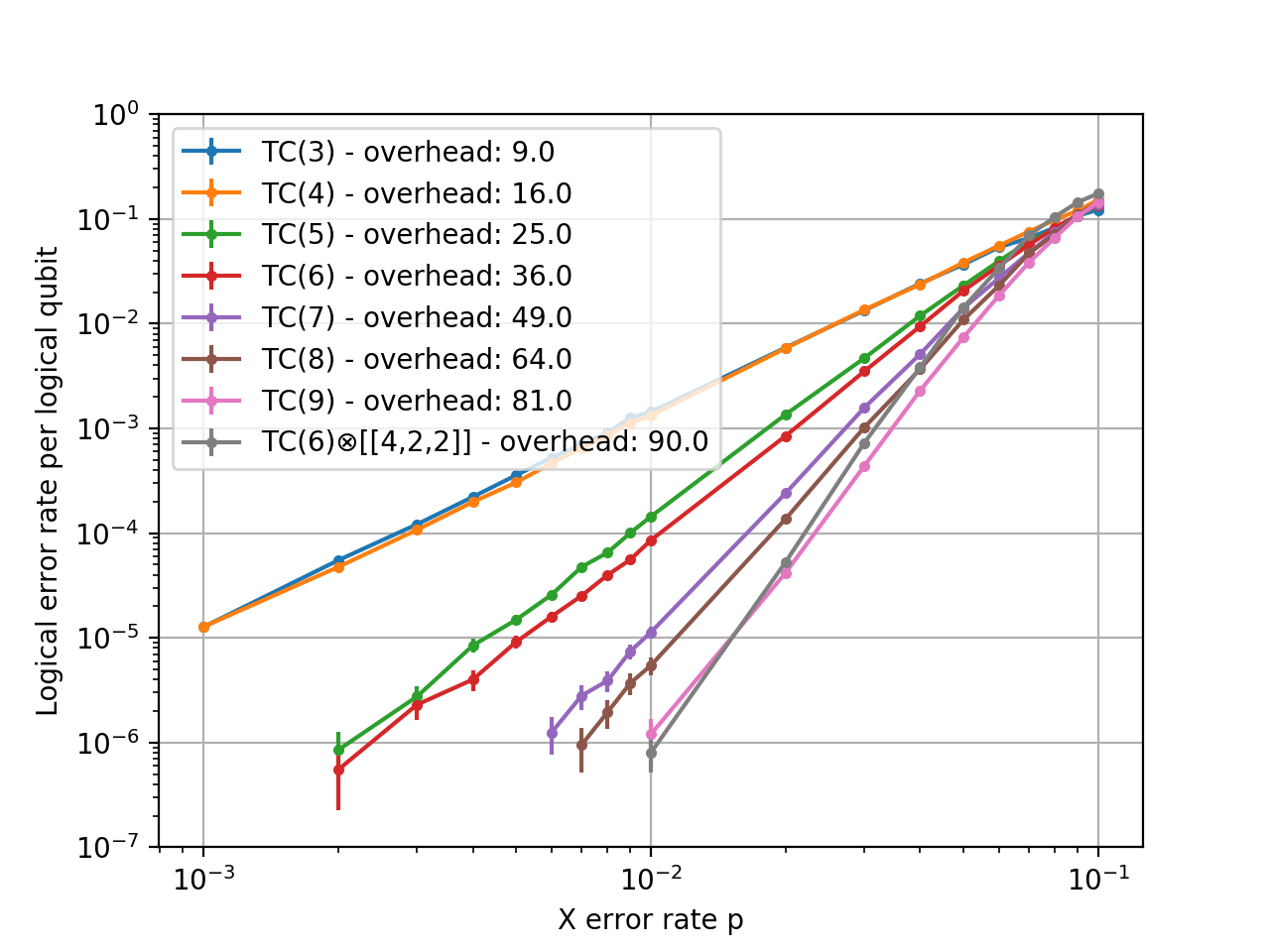}
\caption{Comparison of augmented toric codes $\tc(m) \otimes [[4, 2, 2]]$
with toric codes with smaller or equal overhead for $m=3, 4, 5, 6$.
Toric codes are decoder is the standard union-find decoder and Algorithm~\ref{algo:uf_decoder_modified} is used to decode augmented toric codes.
Augmented toric codes achieve a better logical error rate than toric codes in the regime of low physical error rate.
The overhead reported in these plots is the number of physical qubit per logical qubit.
}
\label{fig:atc_vs_tc}
\end{figure}

In Figure~\ref{fig:atc_vs_tc}, we compare the product $\tc(m) \otimes [[4, 2, 2]]$ with all the toric codes $\tc(\ell)$ achieving a smaller or equal qubit overhead.
Our numerical results show that augmented toric codes with $m \geq 3$ outperform toric codes in the regime of low physical error rate.  At higher physical error rates, the augmented toric codes do not perform as well, and pseudo-thresholds extracted from these figures (i.e., the point at which logical error rate is equal to physical) are worse for the augmented codes.

As an example, the augmented toric code $\tc(6) \otimes [[4, 2, 2]]$ consumes 90 physical qubits per logical qubit and at low physical error rates it achieves a lower logical error rate than the toric code $\tc(9)$ which has comparable overhead.
The origin of the superiority of augmented toric codes is their increased minimum distance which reaches 12 for the code $\tc(6) \otimes [[4, 2, 2]]$, whereas the traditional toric code is limited to distance 9 for the same overhead.

%
%

\subsection{Distance}
The results of \cref{distance} show that the distance of an augmented surface code equals the product of the distances of the large and fixed codes.  Here we show that

\begin{lemma}
If the fixed code has distance $2$, then the union-find decoder (in its simplest form where we do not have separate clusters for each logical operator)  decodes errors with weight less than half the distance.  (It is already known\cite{delfosse2017almost} that for the {\it unaugmented} surface code, the union-find decoder has this property.)

Further, a modification of the algorithm (detailed in the proof of this lemma and not studied elsewhere in the paper) decodes errors with weight less than half the first for any distance of the fixed code.
\begin{proof}
First consider the case where the fixed code has distance $2$.
Suppose an error of weight $w$ occurs.
Let there be $m$
edges in ${\cal E}$ before edge cancellation; these edges have at least one error on them.
So, the sum over diameters of clusters in ${\cal E}$ is bounded by $m$.

After edge cancellation, there may be some edges that have an error on them but are not in ${\cal E}$.  Call this set of edges ${\cal H}$, for ``hidden". There must be at least $2$ errors on each of these edges and so the number of such edges is at most $(w-m)/2$.

Now consider the growth process of clusters.  (This step of the proof is almost exactly the same as the proof\cite{delfosse2017almost} that the union-find decoder decodes up to half the distance for a surface code.)  If a cluster is not valid, then there must be some edge in ${\cal H}$ leaving the cluster.
A single step of the growth will cover half of one of the edges in ${\cal H}$, and will increase the sum of cluster diameters by at most $1$ (it is possible that it joins two clusters and so greatly increases the largest cluster diameter).  
Suppose the growth process terminates after $t$ steps, at which point it has covered at least $t/2$ edges in ${\cal H}$ (hence, $t\leq w-m$) and the
sum of cluster diameters is bounded by $m+t\leq w$.
It is possible at this point that not all edges in ${\cal H}$ are covered: there may be paths in ${\cal H}$ going from
a cluster to itself, and the length of these paths is then bounded by $(w-m)/2-t/2=(w-m-t)/2$.
So,
the sum of the cluster diameters plus the sum of the lengths of these paths is bounded by $m+t+(w-m-t)/2 \leq w$.
So, the algorithm replaces the actual error with clusters of erasures plus possibly some operator which commutes with the stabilizers (i.e., due to paths in ${\cal H}$ which are not covered), with these errors supported on sets of diameter at most $w$. 
If $w$ is smaller than the surface code distance, it decodes correctly.
Remark: we see that the worst case is when all edges in ${\cal H}$ are covered.

Now consider the case where the fixed code has distance $d>2$.  We modify the algorithm as follows.
Each  edge is broken into $d$ subedges.  Note that $d$ may be odd.  Rather than keeping a set of half edges, we keep a set of these subedges.  
Let $w(e)$ denote the error weight on each edge, so that $w=\sum_e w(e)$.

We first perform the edge cancellation step.  Let $c(e)$ denote the weight of the correction applied on edge $e$ if that correction weight is $\leq d/2$; if the correction weight is $>d/2$ then $c(e)=d/2$.  
If an error of weight $w(e)$ occurs, with $w(e)\geq d/2$, then $c(e)\geq d-w(e)$.
We use a minimum weight decoder so that $c(e) \leq w(e)$.  If we fix an error on an edge $e$ with a correction of some weight $c(e)$, we add the $c(e)$ closest subedges to each vertex attached to that edge to the subedge set, so that the total number of subedges on that edge in the subedge set is $2c(e)$.  Thus, in the case that the distance is $2$, any correction means adding both half edges, but for larger distance, it is possible that only part of each edge is added.

We grow so that subedges are added starting closest to the vertex and moving outwards.  Note that if $d$ is odd, it is possible that two clusters are separated by an odd number (for example, $1$) subedge; to deal with this, we add subedges to clusters sequentially, rather than in parallel, or, alternatively, one may further subdivide each subedge into two subsubedges to run growth in parallel.

Let $s(e)$ denote the total number of subedges on edge $e$ which are in the subedge set.  Thus, initially $s(e)=2c(e)$, but $s(e)$ may increase as the algorithm runs.
The sum of cluster diameters initially is $\leq (2/d)\sum_e c(e)$

After applying edge cancellation, the error applied on each edge may be a nontrivial logical operator.
Let ${\cal L}$ denote the set of such edges with a nontrivial logical operator, and for an edge $e\in {\cal L}$, let $h(e)=d-s(e)$ denote the ``hidden weight" on that edge.
Let ${\cal H}$ denote the set of edges $e$ with $h(e)>0$: these are edges in ${\cal L}$ on which not all subedges are in the subedge set.  Let the total hidden weight $h=\sum_{e\in {\cal H}} h(e)$.

Before any growth, immediately
after edge cancellation, the hidden weight is bounded by $h_0 \equiv \sum_{e \, {\rm s.t.} \, w(e)\geq d/2} (d-2c(e))$.
This hidden weight $h$ reduces as a result of the growth process.
If a cluster is not valid, then there must be some edge in ${\cal H}$ leaving the cluster.    So, a single step of step of growth (growing on any single subedge) will reduce the hidden weight by at least $1$ and will increase the sum of cluster diameters by at most $2/d$.

So, if the algorithm terminates after $t$ steps, the sum of cluster diameters is upper bounded by $(2/d)(\sum_e c(e)+t)$,
and $t \leq h_0$.
There may possibly be some hidden weight that remains at the end of the algorithm: this is due to edges with hidden weight $d$
forming paths from a cluster to itself.  Thus, the sum of cluster diameters plus the length of these paths
is bounded by $(2/d)(\sum_e c(e)+t)+(h_0-t)/d$.  This is maximized at $t=h_0$, so this is bounded by $(2/d) \Bigl(\sum_e c(e)+h_0\Bigr)= (2/d) \Bigl( \sum_{e \, {\rm s.t.} \, w(e)<d/2} c(e)+\sum_{e \, {\rm s.t.} \, w(e) \geq d/2} (d-c(e))
\Bigr).$
Note that $c(e)\leq w(e)$ for all $e$ and also $d-c(e)\leq w(e)$ for all $w(e)\geq d/2$.
Hence, the sum of cluster diameters plus path lengths is bounded by $(2/d) \sum_e w(e)=(2/d) w$.
\end{proof}
\end{lemma}

\section{Union-Find Decoder in General}
\label{general}
We now consider how to construct a decoder for a more general choices of large and fixed codes.
As before, we assume that the large code has a union-find decoder, and we construct a union-find decoder for the homological product.

The initialization and growth aspects of the decoder can be chosen in various ways (for example, growing each cluster by a minimal amount).  What we will be concerned with is the routines for testing if a cluster is valid and for decoding a valid cluster.  We will assume that those two routines exist for any cluster of the large code, and we will construct them for the product code for any choice of fixed code in \cref{valid}, \cref{decode}, subject only to the requirement that the fixed code have no redundancies in its stabilizers.

\subsection{Syndrome Validation}
\label{valid}
We consider the question: 
given some set of qubits and some observed syndrome on some set of $X$ checks, is there a $Z$ error pattern on those
qubits which can give rise to that syndrome on those checks, and no errors on other checks?  That is, is the cluster consisting of those qubits and those checks {\it valid}?

Define a {\it sub-cell complex} to be some set of cells of the large code (i.e., some cells of the cellulation if it is a topological code, and some set of $X$- and $Z$-checks and qubits if it is not), such that if a cell is in the subcomplex, all cells in its boundary are also (in the case of a code that is not a topological code, we mean that if some $Z$-stabilizer is in the subcomplex, then all qubits in its support are, and if a qubit is in the subcomplex, then all $X$-stabilizers supported on it are, i.e., given a vector corresponding to some cell of the subcomplex, all cells in the support of the boundary of that vector are in the subcomplex).
We write such a sub-cell complex with regular capital letters such as $P$; corresponding to such a sub-cell complex there is a sub-chain complex which we write with calligraphic letters such as $\cQ$; this is a set of vector spaces $\cQ_j$ corresponding to $j$-cells of the sub-cell complex $P$, and we define the obvious boundary operator on the sub-chain complex.  For brevity, we will refer to both $P$ and $\cQ$ as subcomplexes.

Then, the question at the start of this section is equivalent to the question: given some subcomplex $P$
(such that the given set of qubits is the set of $(q+1)$-cells in $P$, and  such that the given set of checks is the set of $q$-cells in $P$), is there a vector supported on $\cQ_q$ whose boundary 
 is the given syndrome?  Note that since we consider a subcomplex, we are guaranteed that a $Z$ error on a qubit in $P$ will not produce an error on any stabilizer which is not in the set of $(q-1)$-cells of the complex.

Using the language of homology, asking if a syndrome vector can be written as a boundary is equivalent to asking: is the syndrome a closed chain (i.e., does its boundary vanish) and is the syndrome homologically trivial?
This question can be answered in a simple way in one particular important case: that the subcomplex $\cQ$ is a homological product of some subcomplex $\cM$  in the large code with some subcomplex $\cCT$ in the fixed code.  
We will write $M$ to denote the sub-cell complex corresponding to $\cM$.

If $\cC$ is the $[[4,2,2]]$ erasure code, it makes sense to take $\cCT=\cC$ always, but if $\cC$ is a larger code it may be useful in practice to take other choices for $\cCT$ and to ``grow" the subcomplex $\cCT$ in different clusters.
The subcomplex $\cCT$ itself defines some CSS code $\CT$ with some number $\kCT$ of logical qubits.
Note that $\cCT$ has trivial second homology since $\cC$ does.
We will further assume that $\cCT$ has trivial zeroth homology; this holds if $\cCT=\cC$ but may not hold in general.

The vector spaces $\cQ_j$ can be regarded as subspaces of $\cA_j$: they are the subspaces where vectors vanish on entries which do not correspond to cells of $M$.
A syndrome on $M$ is some element of $\cA_q$ which is in this subspace $\cQ_j$.

Then, we claim that 
the following algorithm will determine,
 given some syndrome $v$ in $\cQ_q$, whether or not it is the boundary of some element $w$ of $\cQ_{q+1}$. 
First (this step is done offline, before running the union-find decoder), for each $j=1,\ldots,\kCT$, construct a vector $x_j$ in $\Z_2^{n_C}$ so that these vectors span all possible logical $X$-operators for code $\CT$.  In the terminology of topology, the cohomology classes $[x_j]$ must be independent.
Then, for each 
 $j=1,\ldots,\kCT$, define a vector $\tilde v_j$ in $\cM_{q-1}$ such that, for each $(q-1)$-cell $e$ of $M$, the coefficient of $\tilde v_j$
 on that cell is given by the inner product of the vector of coefficients of $v$ on $e$ with $x_j$.
 We call this vector $\tilde v_j$ the {\it partial inner product} of $v$ with $x_j$ and we write it $v_j=(v,x_j)$ in an abuse of notation.
Then, $v$ is the boundary of some $w$ if and only if $\partial_v=0$ and each $\tilde v_j$ is the boundary of some vector in $\cM_j$.  The question of whether $\tilde v_j$ is such a boundary, however, is precisely the question that a union-find decoder for the large code must solve so by assumption we have an algorithm to do this, and we will show that the question of whether $\partial v=0$ can always be solved in linear time.

Let us show that this algorithm is correct.
\begin{lemma}
Let $\cQ$ be a product $\cM \otimes \cCT$ where $\cM,\cCT$ are subcomplexes of the large code and fixed code, respectively.
Assume that $\cCT$ has trivial zeroth and second homology.
Consider some syndrome $v\in \cA_q$ such that $v$ is supported on $\cQ_q$.

Then $M$ is a valid cluster for $v$, meaning that $v$ is a boundary of some vector in $\cQ_{q+1}$, iff
$\partial_\cQ v = 0$ and, for all $j \in \{1,\ldots,\kCT\}$, the vector $\tilde v_j$ is a boundary in $\cM$, where $\tilde v_j$ is the partial inner product $\tilde v_j=(v,x_j)$.

Note that $\tilde v_j$ is a boundary in $\cM$ iff its inner product with all $(q-1)$-th cohomology representatives of $\cM$ vanishes; in the case that the large code is the surface code, there is one such representative and this gives precisely the $j$-th entry of of the validity vector considered previously.
\begin{proof}
$v$ is a boundary (i.e., it represents trivial homology) iff $\partial v=0$ and its inner product
with 
a basis of cohomology representatives vanishes.  (This is a consequence of the universal coefficient theorem though it can be proven more simply in the case of ${\mathbb Z}_2$ coefficients.)

By K\"{u}nneth, given that $\cCT$ has trivial zeroth and second homology,
 a basis $x_k$ for those cohomology representatives can be given by the product of nontrivial $(q-1)$-th cohomology representatives of $\cM$ with first cohomology representatives of $\cCT$, i.e., logical $X$ operators, namely the $x_j$.
Verifying that all these inner products vanish is equivalent to verifying, that for each $j=1,\ldots,\kCT$,
the inner product of $\tilde v_j$ with all cohomology representatives of $\cM$ vanishes, which in turn is equivalent to requiring that $\tilde v_j$ be a boundary in $\cM$.
\end{proof}
\end{lemma}

Now consider how to efficiently check that $\partial v=0$.  
The vector $\partial v$ is in the space
$(\cM_{q-3} \otimes \cCT_2) \oplus (\cM_{q-2} \otimes \cCT_1) \oplus (\cM_{q-1} \otimes \cCT_0).$
This can clearly be done efficiently if we have an efficient representation of the boundary operator $\partial$, simply by checking it for each $(q-3)$-,$(q-2)$-, or $(q-1)$-cell of $M$.
Further, in a union-find decoder there is no need to check this constraint $\partial v=0$ on cells $e$ in the interior of $M$ (meaning those cells that are not attached to any cell in $L\setminus M$), as the constraint is automatically satisfied on those cells by the assumption that the observed syndrome is the boundary of {\it some} error pattern on $L$.
So, one simply needs to check this constraint on cells at the boundary of $M$; in a union find decoder, this constraint then only needs to be checked once for each cell in $M$, when the cell is on the boundary.

\subsection{Decoding}
\label{decode}
We have given an algorithm to determine if some vector $v\in \cQ_q$ is a boundary of some $w$.  We now give an algorithm that, if the answer to the previous question is yes, will find such a $w$.  As before, we assume that we have an algorithm to solve this problem for the large code.

Recall that $v\in (\cM_{q-2} \otimes \cCT_2) \oplus (\cM_{q-1}\otimes \cCT_1) \oplus (\cM_q \otimes \cCT_0)$ and we wish to find
a $w\in (\cM_{q-1} \otimes \cCT_2) \oplus (\cM_{q}\otimes \cCT_1) \oplus (\cM_{q+1} \otimes \cCT_0)$ with $v=\partial w$.
We will in fact find a $w$ which vanishes in the subspace $\cM_{q+1} \otimes \cCT_0$.

The construction is slightly notationally laborious in general but is straightforward: we add different boundaries to $v$ to cancel various components of it on subspaces $\cM_q \otimes \cCT_0$, $\cM_{q-1} \otimes \cCT_1$, and $\cM_{q-2} \otimes \cCT_2$ in turn.
The first cancellation and the last cancellation are straightforward using the vanishing of zeroth and second homology of $\cCT$ (in fact the last cancellation happens ``automatically" as a result of a previous cancellation).  The second cancellation is a little trickier due to nontrivial first homology of $\cCT$, but is reduced to the problem for the large code.

We begin with the first cancellation; this step is precisely the edge cancellation step of \cref{sddecode}.
Let $v_0$ be the component of $v$ in subspace $\cM_q \otimes \cCT_0$.
Let $\partial_M$ be the boundary operator on $\cM$.  Define $``\pCT^{-1}"$ to be an operator such that
$\pCT ``\pCT^{-1}" y=y$ for any vector $y\in \cCT_0$; in words, given an syndrome for $\CT$, $``\pCT^{-1}"$ computes an error pattern that gives that observed syndrome; this operator is precisely the decoder $D_Z$ used in \cref{sddecode}.  Such an operator exists because the zeroth homology of $\CT$ is trivial.
Let $w_0=``\pCT^{-1}" v_0$.
Then, $\partial w_0=v_0+\partial_M ``\pCT^{-1} v_0".$
So, the sum $v+\partial w_0$ vanishes in subspace $\cM_q \otimes \cCT_0 $.

Replace $v$ with $v+\partial w_0$.  We have reduced to the problem of finding $w$ such that $v=\partial w$ for $v$ vanishing in subspace $\cM_q \otimes \cCT_0$.

Now, take vector $v$ and, for each $j=1,\ldots,\kCT$, compute a vector $\tilde v_j$ as above.  Then, apply the union-find decoder for the large code to find a $\tilde w_j$ such that $\partial_B \tilde w_j=\tilde v_j$.  For each of the $X$ logical operators $x_j$ find corresponding $Z$ logical operators $z_j$, so that $(z_j,x_k)=\delta_{j,k}$.  Then, consider vector
$v'=v+\partial(\sum_j \tilde w_j \otimes z_j).$
Since $z_j$ is a $Z$ logical operator, $\pCT z_j=0$, so $v'$ still vanishes in subspace $\cM_q \otimes \cCT_0$.
Let $v'_1$ be the component of $v'$ in subspace $\cM_{q-1} \otimes \cCT_1$.
Since $\partial v'=0$, we have that for every $(q-1)$-cell, the boundary (using $\pCT$) of the vector of coefficients of $v'$ on that cell vanishes, i.e., that vector of coefficients is a cycle.  However, the term $\partial(\sum_j \tilde w_j \otimes z_j)$ in the sum for $v'$ guarantees then that that vector represents trivial homology.  So, it is a boundary.
So, for each $(q-1)$-cell $e$, there is some vector $w_e$ such that $\partial_e w_e$ gives the vector of coefficients of $v'$ on cell $e$.

So, consider vector $v''=v'+\partial(\sum_e 1_e\otimes w_e),$ where $1_e$ denotes the basis vector corresponding to cell $e$.
Then, $v''$ vanishes on $(\cM_{q} \otimes \cCT_0 )\oplus (\cM_{q-1} \otimes     \cCT_1 )$.
So, the only nonvanishing component of $v''$ is on $\cM_{q-2} \otimes \cCT_2 $.  Then, since $\partial v''=0$, by computing $\partial v''$ projected onto subspace $\cM_{q-2} \otimes \cCT_{1}$ and using that the second homology of $\cCT$ is trivial so that the only vector in $\cCT_2$ with vanishing boundary is the zero vector, we find that $v''=0$.

\section{Distance}
\label{distance}
The distance $d$ of a homological product obeys the upper bound $d\leq d_B d_C$ by the K\"{u}nneth formula.
It is known\cite{bravyi2014homological} that this bound is not necessarily tight.  However, we now show that in many cases, including many choices of a topological large code, this bound indeed is tight.

We have the following trivial lower bound for the distance $d_B$ of the large code: if some equivalence class $[x]$ for $q$-th cohomology has at least $m$ distinct representatives, $x_1,\ldots,x_m$, such that the support of $x_i$ is disjoint from the support of $x_j$ for $i\neq j$, then any representative of an equivalence class $[z]$ for $q$-th homology which has nontrivial inner product with $[x]$ will have weight at least $m$, since, of course, each representative must have some intersection with each $x_i$.  If this holds for all equivalence classes of $q$-th cohomology, then the $Z$ distance of the code is at least $m$.

As an example of this, consider a toric code on a torus; assume the torus is $L$-by-$L$ and call the two directions ``vertical" and ``horizontal".  One choice of logical $Z$ operator is a string of Pauli $Z$ running in the vertical direction.  There are $L$ different such strings of minimal length, corresponding to different locations of the string in the horizontal direction.  So, any logical $X$ operator which anti-commutes with it must have weight at least $L$.

If some equivalence class $[x]$ has this property of having $m$ distinct representatives with disjoint support, we say that class has property $(*)$.

This bound has a trivial extension (we use $(q+1)$-th rather than $q$-th homology now since we intend to apply this bound to the product code, but of course this bound is true for any $q$):
 if some equivalence class $[x]$ for $(q+1)$-th cohomology has at least $m$ distinct representatives, $x_1,\ldots,x_m$, such that the support of $x_i$ is disjoint from the support of $x_j$ for $i\neq j$,
 and such that any representative
 of
an equivalence class $[z]$ for $(q+1)$-th homology which has nontrivial inner product with $x_i$ for any $i$ must have weight at least $d_C$ on the support of $x_i$, then any representative of $[z]$ must have weight at least $md_C$.

If some class $[x]$ has this
 this property of having $m$ distinct representatives with disjoint support and with lower bound $d_C$ on the weight of the intersection, we say that this class has property $(**)$.

We now show that if the class $[x]$ for the large code has property $(*)$, then the class $[x \otimes \ell]$ has property $(**)$ if $\ell$ represents an $X$ logical operator of fixed code $C$.  Proof: let $z$ be some representative of a class with nontrivial inner product with $[x\otimes \ell]$.  Fix some $i$ to choose a representative $x_i$ of $[x]$.  
Let us say, given a vector $v$ in some subspace $\cB_{q}\otimes \cC_1$ or $\cB_q \otimes \cC_0$ that the partial inner product of that vector with some other vector $u$ in $\cB_q$ is a vector in either $\cC_1$ or $\cC_0$, respectively, defined in the obvious way: regard $v$
as being a vector (of length ${\rm dim}(\cC_1)$ or ${\rm dim}(\cC_0)$, respectively) of vectors in $\cB_q$, and take the inner product of each such vector with $u$.
Consider the projection of $z$ onto subspace
$\cB_{q}\otimes \cC_1$ and take its partial inner product with $x_i$ to get a vector $u$ in $\Z_2^{n_C}$.  We have $(u,\ell)=(z,x\otimes \ell)=1$.  We will next show that $\partial u=0$.  This will show that $u$ represents a $Z$ logical operator for $C$ and so has weight at least $d_C$, proving the desired result.

To show $\partial u=0$, consider the projection of $\partial z$ onto subspace $\cB_q \otimes \cC_0$; call this $s$.  Of course, since $\partial z=0$ we have $s=0$.
Take the partial inner product of $s$ with $x_i$ to get
a vector $t$; again $t=0$.
However, $t$ is the sum of two contributions; the first is the partial inner product with $x_i$ of the boundary of the projection of $z$ into $\cB_{q+1}\otimes \cC_0$, but this term vanishes since $x_i$ is a cocycle and so has vanishing inner product with any boundary. The second term is exactly equal to $\partial u$, so indeed $\partial u=0$.

Then, with these results, for many cases such as the large code being a toric code on a torus or more generally any topological code on a torus, the distance of the homological product code equals the product $d_B d_C$.  We leave open the question of the distance of the product code when the large code is a topological code with some boundaries.

\bibliography{hp-ref}

\begin{thebibliography}{10}

\bibitem{freedman2013quantum}
Michael~H Freedman and Matthew~B Hastings.
\newblock Quantum systems on non-$ k $-hyperfinite complexes: A generalization
  of classical statistical mechanics on expander graphs.
\newblock {\em QIC}, 14:144, 2014.

\bibitem{bravyi2014homological}
Sergey Bravyi and Matthew~B Hastings.
\newblock Homological product codes.
\newblock In {\em Proceedings of the forty-sixth annual ACM symposium on Theory
  of computing}, pages 273--282, 2014.

\bibitem{hastings2017weight}
Mathew~B Hastings.
\newblock Weight reduction for quantum codes.
\newblock {\em Quantum Information \& Computation}, 17(15-16):1307--1334, 2017.

\bibitem{evra2020decodable}
Shai Evra, Tali Kaufman, and Gilles Z{\'e}mor.
\newblock Decodable quantum ldpc codes beyond the $\sqrt{n}$ distance barrier
  using high dimensional expanders.
\newblock {\em arXiv preprint arXiv:2004.07935}, 2020.

\bibitem{audoux2015tensor}
Benjamin Audoux and Alain Couvreur.
\newblock On tensor products of css codes.
\newblock {\em arXiv preprint arXiv:1512.07081}, 2015.

\bibitem{tillich2013quantum}
Jean-Pierre Tillich and Gilles Z{\'e}mor.
\newblock Quantum ldpc codes with positive rate and minimum distance
  proportional to the square root of the blocklength.
\newblock {\em IEEE Transactions on Information Theory}, 60(2):1193--1202,
  2013.
\newblock \href {https://doi.org/10.1109/isit.2009.5205648}
  {\path{doi:10.1109/isit.2009.5205648}}.

\bibitem{leverrier2015quantum}
Anthony Leverrier, Jean-Pierre Tillich, and Gilles Z{\'e}mor.
\newblock Quantum expander codes.
\newblock In {\em 2015 IEEE 56th Annual Symposium on Foundations of Computer
  Science}, pages 810--824. IEEE, 2015.
\newblock \href {https://doi.org/10.1109/focs.2015.55}
  {\path{doi:10.1109/focs.2015.55}}.

\bibitem{Fawzi_2018}
Omar Fawzi, Antoine Grospellier, and Anthony Leverrier.
\newblock Constant overhead quantum fault-tolerance with quantum expander
  codes.
\newblock In {\em 2018 {IEEE} 59th Annual Symposium on Foundations of Computer
  Science ({FOCS})}. {IEEE}, oct 2018.
\newblock URL: \url{https://doi.org/10.1109%2Ffocs.2018.00076}, \href
  {https://doi.org/10.1109/focs.2018.00076}
  {\path{doi:10.1109/focs.2018.00076}}.

\bibitem{panteleev2019degenerate}
Pavel Panteleev and Gleb Kalachev.
\newblock Degenerate quantum ldpc codes with good finite length performance.
\newblock {\em arXiv preprint arXiv:1904.02703}, 2019.

\bibitem{quintavalle2020single}
Armanda~O Quintavalle, Michael Vasmer, Joschka Roffe, and Earl~T Campbell.
\newblock Single-shot error correction of three-dimensional homological product
  codes.
\newblock {\em arXiv preprint arXiv:2009.11790}, 2020.

\bibitem{delfosse2017almost}
Nicolas Delfosse and Naomi~H Nickerson.
\newblock Almost-linear time decoding algorithm for topological codes.
\newblock {\em arXiv preprint arXiv:1709.06218}, 2017.

\bibitem{vrana2019homological}
P{\'e}ter Vrana and M{\'a}t{\'e} Farkas.
\newblock Homological codes and abelian anyons.
\newblock {\em Reviews in Mathematical Physics}, 31(10):1950038, 2019.
\newblock \href {https://doi.org/10.1142/s0129055x19500387}
  {\path{doi:10.1142/s0129055x19500387}}.

\bibitem{criger2016noise}
Ben Criger and Barbara Terhal.
\newblock Noise thresholds for the [[4, 2, 2]]-concatenated toric code.
\newblock {\em arXiv preprint arXiv:1604.04062}, 2016.

\bibitem{kitaev2003fault}
A~Yu Kitaev.
\newblock Fault-tolerant quantum computation by anyons.
\newblock {\em Annals of Physics}, 303(1):2--30, 2003.
\newblock \href {https://doi.org/10.1016/s0003-4916(02)00018-0}
  {\path{doi:10.1016/s0003-4916(02)00018-0}}.

\bibitem{freedman2001projective}
Michael~H Freedman and David~A Meyer.
\newblock Projective plane and planar quantum codes.
\newblock {\em Foundations of Computational Mathematics}, 1(3):325--332, 2001.
\newblock \href {https://doi.org/10.1007/s102080010013}
  {\path{doi:10.1007/s102080010013}}.

\bibitem{bombin2007homological}
Hector Bombin and Miguel~A Martin-Delgado.
\newblock Homological error correction: Classical and quantum codes.
\newblock {\em Journal of mathematical physics}, 48(5):052105, 2007.
\newblock \href {https://doi.org/10.1063/1.2731356}
  {\path{doi:10.1063/1.2731356}}.

\bibitem{delfosse2020peeling}
Nicolas Delfosse and Gilles Z{\'e}mor.
\newblock Linear-time maximum likelihood decoding of surface codes over the
  quantum erasure channel.
\newblock {\em Physical Review Research}, 2(3):033042, 2020.
\newblock \href {https://doi.org/10.1103/physrevresearch.2.033042}
  {\path{doi:10.1103/physrevresearch.2.033042}}.

\end{thebibliography}

\appendix
\section{Union-Find decoder with modified growth}
\label{app:UF_modified}

In this section, we describe a version of Algorithm~\ref{algo:uf_decoder} with 
a different growth strategy. 

Instead of growing a cluster until its validity vector becomes trivial, we can consider independently each component of the validity vector. This produces smaller clusters, increasing the chances to keep the clusters correctable.

For each logical $x_i$, we perform a growth step and a peeling step.
For the growth step of $x_i$, a cluster $\cluster$ is said to be valid if
$
\val_i(\cluster) = \sum_{v \in \cluster} (s(v) | x_i) = 0 \pmod 2.
$
The growth step of $x_i$ produces clusters such that the $i$ th component of the validity vector is trivial.
Then, the peeling step of $x_i$ is the logical part of the peeling step of Algorithm~\ref{algo:uf_decoder}.

Numerically, we observe a slight improvement of the performance in comparison with Algorithm~\ref{algo:uf_decoder}.
Our numerical simulations are based on this variant of the decoder. 

\begin{algorithm}
\caption{Union-Find decoder for augmented surface codes - Version 2}
\label{algo:uf_decoder_modified}

\begin{algorithmic}[1]
\REQUIRE The syndrome $s(x)$ as in \eqref{eqn:syndrome} of an error $x \in A_2$.\\
\ENSURE An estimation $\tilde x \in A_2$ of $x$.

\STATE Set $\tilde x = 0$ and ${\cal E} = \{ e \in E \ | \ s(e) \neq 0 \}$.
\STATE {\bf Edge cancellation:} 
\STATE Run over all edges $e = \{ u, v \} \in E$ and do:
\STATE \hspace{1cm} Compute $\tilde x(e) = D_X(s(e))$.
\STATE \hspace{1cm} Add $e \otimes \tilde x(e)$ to $\tilde x$.
\STATE \hspace{1cm} Add $u \otimes s(e)$ and $v \otimes s(e)$ to $s$.
\STATE For $i=1, \dots, k_C$ do:
\STATE \hspace{1cm} {\bf Logical growth:} 
\STATE \hspace{1cm} Initialize clusters with a single vertex $\cluster_{v} = \{v\}$.
\STATE \hspace{1cm} Merge clusters connected by an edge of ${\cal E}$.
\STATE \hspace{1cm} While there exists at least one cluster with $\val_i(\cluster) \neq 1$ do:
\STATE \hspace{2cm} Select a cluster $\cluster$ with $\val_i(\cluster) \neq 0$ with minimum boundary.
\STATE \hspace{2cm} Grow $\cluster$ by one half-edge.
\STATE \hspace{2cm} Update $\val_i(\cluster)$.
\STATE \hspace{1cm} Set ${\cal E}_i'$ to be the set of all edges fully covered by the grown clusters.
\STATE \hspace{1cm} {\bf Logical peeling:} 
\STATE \hspace{1cm} Construct a spanning forest ${\cal F}$ of the subgraph ${\cal E}_i'$ of the surface.
\STATE \hspace{1cm} While ${\cal F} \neq \emptyset$ do:
\STATE \hspace{2cm} Select an edge $e = \{ u, v \}$ of ${\cal F}$ such that $u$ is a leaf.
\STATE \hspace{2cm} If $(s(u)|x_i) = 1$ do:
\STATE \hspace{3cm} Add $e \otimes z_i$ to $\tilde x$.
\STATE \hspace{3cm} Add $u \otimes z_i$ and $v \otimes z_i$ to $s$.
\STATE \hspace{2cm} Remove $e$ from ${\cal F}$
\STATE {\bf Non-Logical Peeling:} 
\STATE Construct a spanning forest ${\cal F}$ for the initial set ${\cal E}$.
\STATE While ${\cal F} \neq \emptyset$ do:
\STATE \hspace{1cm} Select an edge $e = \{ u, v \}$ of ${\cal F}$ such that $u$ is a leaf.
\STATE \hspace{1cm} Add $e \otimes s(u)$ to $\tilde x$.
\STATE \hspace{1cm} Add $u \otimes s(u)$ and $v \otimes s(u)$ to $s$.
\STATE \hspace{1cm} Remove $e$ from ${\cal F}$
\STATE {\bf Residual Node Correction:} 
\STATE For all $v \in V$ do:
\STATE \hspace{1cm} Compute $\tilde x(v) = D_Z^T(s(v))$.
\STATE \hspace{1cm} Add $v \otimes \tilde x(v)$ to $\tilde x$.
\STATE Return $\tilde x$.
\end{algorithmic}
\end{algorithm}

\end{document}